\documentclass[mathleft
]{an}
\usepackage{graphicx}
\usepackage{times}
\usepackage{bm}
\usepackage{color}
\begin{document}
\sloppy

\Pagespan{789}{}
\Yearpublication{2006}%
\Yearsubmission{2005}%
\Month{11}%
\Volume{999}%
\Issue{88}%

\newcommand{\EQ}{\begin{equation}}
\newcommand{\EN}{\end{equation}}
\newcommand{\eq}[1]{(\ref{#1})}
\newcommand{\Eq}[1]{equation~(\ref{#1})}
\newcommand{\Eqs}[2]{equation~(\ref{#1}) and (\ref{#2})}
\newcommand{\Fig}[1]{Fig.~\ref{#1}}
\newcommand{\Figs}[2]{Figs.~\ref{#1} and \ref{#2}}
\newcommand{\Tab}[1]{Table~\ref{#1}}
\newcommand{\bra}[1]{\langle #1\rangle}
\hyphenation{wave-num-ber Abramo-wicz Rey-nolds}


\def\onethird{{\textstyle{1\over3}}}
\def\onehalf{{\textstyle{1\over2}}}

\newcommand{\yprl}[3]{ #1, {PRL,} {#2}, #3}

\title{Turbulent stresses as a function of shear rate in a local disk model}

\author{A. J. Liljestr\"om\inst{1}\fnmsep\thanks{Corresponding author:
  \email{anne.liljestrom@helsinki.fi}\newline}
\and M. J. Korpi\inst{1}
\and P. J. K\"apyl\"a\inst{1}
\and A. Brandenburg\inst{2}
\and W. Lyra\inst{3}
}
\titlerunning{Turbulent stresses in a local disk model}
\authorrunning{Liljestr\"om et al.}
\institute{
Observatory, University of Helsinki, PO BOX 14, FI-00014 University of Helsinki, Finland
\and 
NORDITA, AlbaNova University Center, Roslagstullsbacken
              23, SE-10691 Stockholm, Sweden
\and
Department of Physics and Astronomy, Uppsala Astronomical 
Observatory, Box 515, 751 20 Uppsala, Sweden}

\received{15 Jul 2008}
\accepted{11 Nov 2008}


\abstract{%
We present local numerical models of accretion disk turbulence driven by 
the magnetorotational instability with varying shear rate. The resulting 
turbulent stresses are compared with predictions of a closure model in 
which triple correlations are modelled in terms of quadratic correlations. 
This local model uses five nondimensional parameters to describe the 
properties of the flow. We attempt to determine these closure parameters 
for our simulations and find that the model does produce qualitatively 
correct behaviour. In addition, we present results concerning the shear 
rate dependency of the magnetic to kinetic energy ratio. We find both the 
turbulent stress ratio and the total stress to be strongly dependent on 
the shear rate.}

\maketitle

\section{Introduction}

Since the work of Balbus \& Hawley (1991) it is now generally accepted 
that turbulence in accretion disks is caused by the linear 
magneto-rotational instability (hereafter MRI) that was originally 
discovered by Velikhov (1959) in connection with Couette flow of liquid 
metals. This linear instability can be excited in sufficiently ionized, 
differentially rotating fluids where the angular momentum decreases with 
increasing radius. When the fluid is threaded by a magnetic field, 
differential rotation causes stretching of the field lines. The tension 
that builds up opposes the shearing, acting to enforce rigid rotation. If 
the field is subthermal, the interplay between differential rotation and 
magnetic tension destabilizes the fluid, resulting in linear growth of 
Reynolds and Maxwell stresses, which transport angular momentum outwards. 
The instability eventually leads to a fully turbulent, non-linear state.

Let us assume a rotation profile of the form $\Omega \propto r^{-q}$, 
where $\Omega$ is the angular velocity at distance $r$. A necessary 
condition for the MRI to be excited is $q>0$, i.e.\ the angular velocity 
decreases outward. The Keplerian case with $q=1.5$ has been extensively 
studied by means of numerical simulations making use of the shearing box 
approximation (e.g., Hawley, Gammie \& Balbus 1995, 1996; Brandenburg et al. 
1995; Johansen \& Klahr 2005), as well as in global disks (e.g.\ Armitage 
1998; Hawley 2001; Lyra et al. 2008). These studies have shown that the 
MRI leads to a fully turbulent saturated state in which the Maxwell stress 
is responsible for the majority (about 80\%) of the angular momentum 
transport. Even in the absence of vertical density stratification a 
turbulent small-scale magnetic field can be maintained by dynamo action 
(Hawley et al. 1996). When stratification is present, cyclic large-scale 
dynamo action can be excited (Brandenburg et al. 1995).

The dependence of Reynolds and Maxwell stresses on the shear rate has been 
investigated numerically by Abramowicz, Brandenburg \& Lasota (1996) in 
the presence of stratification and Hawley, Balbus \& Winters (1999) in the absence of 
stratification. It turns out that near $q = 0$ the Reynolds stress, which 
couples to the large-scale vorticity $W = (2 - q)\Omega$, is small due to 
the strong stabilizing effect of the vorticity (Hawley et al. 1999). This 
makes the Maxwell to Reynolds stress ratio very high. As the shear rate 
$q\Omega$ increases, the shear-coupled Maxwell and Reynolds stresses 
increase due to the decrease of the vorticity. However, the growth of the 
Reynolds stress is significantly faster, so the stress ratio diminishes 
with increasing $q$.

The interest in MRI-generated turbulent stresses as a function of $q$ was 
rekindled in a recent study where a linear analysis of the Reynolds and 
Maxwell stresses was presented (Pessah, Chan \& Psaltis 2006a, hereafter PCP06). 
They derived a simple relation for the stress ratio, depending only on the 
shear parameter $q$. Comparing this result with the non-linear simulations 
of Hawley et al. (1999), they find that, even in the saturated turbulent 
regime, the stress ratio does indeed depend almost entirely on $q$ alone, 
and that there is only a weak dependence on other properties of the flow 
or on the initial conditions.

The Shakura-Sunyaev viscosity parameter $\alpha$ (Shakura \& Sunyaev 1973) 
is still a popular tool to link accretion disk observations to theory. 
The so-called $\alpha$ model of disks is based on the assumption that the 
turbulent stresses $T_{r\phi}$ scale linearly with the thermal pressure; 
$\alpha \le 1$ being the proportionality factor.
In this formalism, the outward transport of angular momentum is 
characterized by a turbulent viscosity of the form \begin{equation} 
\label{eq:ass}
  \nu_{\rm t} = \alpha c_{\rm s} H,
\end{equation}
where the eddy viscosity is assumed to scale with the sound speed $c_{\rm s}$
and the correlation length with the disk scale height $H$.

As this model linearizes the turbulent system into a standard 
Navier-Stokes fluid suitable for analytical manipulation, it has been an 
invaluable tool in developing theory of accretion disk dynamics, even 
though the parametrization offers no explanation of what is causing the 
viscosity. Even after the realization of the astrophysical importance and 
subsequent reinvigorated interest in the theory of the MRI, the 
$\alpha$-model is still widely used.

New, more physically motivated models have been developed recently to 
describe the angular momentum transport in accretion disks (e.g.\ Kato \& 
Yoshizawa 1995; Pessah, Chan \& Psaltis 2006b; Ogilvie 2003, hereafter O03). A 
common characteristic of these models is the treatment of the problem of 
angular momentum transport: the governing magnetic and kinetic equations 
are divided into linear and non-linear terms. The non-linear correlation 
functions are then modelled by closing the system of equations with 
approximate expressions that still embody the physics but are considerably 
easier to solve (such closure models are commonly used in modelling 
turbulence; see references in O03). It should be noted that the 
$\alpha$-model is mathematically equivalent to the simplest closure model, 
where the correlation functions are modelled by one single coefficient. 
All these models also operate in the absence of mean magnetic fields. A 
first attempt to validate the O03 model was made by Garaud \& Ogilvie 
(2005) in connection with shear flow where linear and non-linear 
instabilities were found to be well reproduced by the model.

In the present study, the turbulent stresses are extracted as functions of 
$q$ from non-stratified local numerical models with zero net flux using 
the shearing box approximation. The simulation data are compared with the 
linear results of PCP06 and the non-linear closure model of O03. There are 
a few other closure models that describe the behaviour of local, 
magnetohydrodynamic turbulence. In particular, the model of Pessah et al. 
(2006b) assumes a uniform vertical field in the disk which is not 
incorporated in our three-dimensional simulations. Therefore we 
concentrate here mainly on the closure model of O03.

The remainder of the paper is organised as follows: in Sections \ref{lmt}, 
\ref{nlcm}, and \ref{themod} the linear MRI model of Pessah et al. 
(2006b), the non-linear closure model of O03, and the numerical model are 
presented, respectively. Furthermore, the results and related discussion 
are presented in Sections \ref{results} and \ref{discuss}.

\section{Stress ratio in the model of PCP06}
\label{lmt}

Here we briefly summarize the formalism employed in the model of PCP06. 
Using the kinematic hydromagnetic equations, PCP06 calculated the ratio of 
the relevant components of Reynolds and Maxwell stresses in a local 
approximation. We use here a Cartesian frame of reference, where $x$, $y$ 
and $z$ denote the radial, toroidal and vertical directions, respectively. 
The relevant component of the stress tensor is then $T_{xy}$, which can be 
decomposed as
\EQ
T_{xy}\equiv R_{xy}-M_{xy}\;,
\label{Txy}
\EN
where
\EQ
R_{ij}=\bra{\rho u_i u_j}\;,\quad
M_{ij}=\bra{b_i b_j}/\mu_0\;,
\EN
are the Reynolds and Maxwell stresses, respectively, and $\bm{u}$ is the 
departure from the mean flow, $\bm{b}$ is the departure from the mean 
magnetic field, $\rho$ is the density, $\mu_0$ the vacuum permeability, 
and angular brackets denote a suitable volume average. Throughout our 
paper, compressibility effects are ignored in the turbulence models, 
i.e.\ $\rho=\rho_0$ is assumed constant, even through the simulations are 
fully compressible. The rms velocity in our simulations remains 
considerably smaller than unity and the flow is therefore subsonic.

Using linear theory, PCP06 found that for given wavenumber $k$ the stress 
ratio is given by
\EQ\label{eq:kstress}
-\frac{M_{xy}(k)}{R_{xy}(k)}
=1+\frac{2(2-q)\Omega_0^2}{k^2u_{\rm A}^2+\gamma_k^2}\;, 
\EN
where $\Omega_0$ is the angular velocity, $v_{\rm A}=B_0/\sqrt{\mu_0\rho_0}$
is the Alfv\'en speed based on a constant vertical magnetic field, and
$\gamma_k$ is the corresponding growth rate of the mode with wavenumber $k$.
Moreover, for the fastest growing mode with
\EQ\label{eq:kmax}
\gamma_{k{\rm max}}/\Omega_0={\textstyle{1\over2}}q\;,\quad
v_{\rm A}^2k_{\max}^2/\Omega_0^2=q-{\textstyle{1\over4}}q^2\;,
\EN
they find
\EQ\label{eq:linstress}
-\frac{M_{xy}}{R_{xy}} = \frac{4-q}{q}\;.
\EN
This expression shows that, for the relevant case with $q<2$, the 
magnitude of the Maxwell stress is always larger than that of the Reynolds 
stress. This formula provides a strikingly simple prediction of the stress 
ratio which is in good agreement with simulation data, even if there is no 
imposed magnetic field (PCP06). One envisages that the relevant wavenumber 
is able to adjust itself to the value where the growth rate is maximal.

\section{The O03 closure model}
\label{nlcm}

The local closure model of O03 includes the linear interaction of the 
turbulent stress tensors with shear and rotation. The linearized evolution 
equations for the Maxwell and Reynolds stresses can be derived directly 
from the basic MHD equations and are fairly straightforward to solve 
numerically. For modelling the non-linear triple correlations of 
fluctuating quantities and the small-scale diffusion, five dimensionless, 
positive definite coefficients appear in the closed system of equations. A 
closure model is needed to deal with these non-linear terms which are 
described by physical effects and constrained by symmetry properties and 
dimensional considerations.

The five closure coefficients stand for the turbulent dissipation of the 
Reynolds stresses ($C_1$), their isotropization ($C_2$), the effect of the 
small-scale Lorentz-force as a source for $R_{ij}$ combined with a sink 
for $M_{ij}$ ($C_3$), a source due to small-scale dynamo for $M_{ij}$ 
combined with a sink for $R_{ij}$ ($C_4$), and turbulent dissipation of 
the Maxwell stresses ($C_5$).

The evolution equations of this model are given by
\begin{eqnarray}\label{eq:timdepeq}
\partial_t R_{ij}&=&-{\cal L}_{ij}^{{\rm R}}-\tau^{-1}\rho_0^{-1/2}({\cal N}_{ij}^{\rm R} + {\cal I}_{ij}), \\
\label{eq:timdepeq2}
\partial_t M_{ij}&=&{\cal L}_{ij}^{{\rm M}}+\tau^{-1}\rho_0^{-1/2}{\cal N}_{ij}^{\rm M},
\end{eqnarray}
where $\tau=L/U$ is the turnover time, $U=R^{1/2}$ is the rms velocity,
$L$ is the typical scale of the energy-carrying eddies, and
\begin{eqnarray}
R\equiv R_{ii}=\langle\rho(u_x^2+u_y^2+u_z^2)\rangle,
\end{eqnarray}
\begin{eqnarray}
M\equiv M_{ii}=\langle b_x^2+b_y^2+b_z^2\rangle/\mu_0,
\end{eqnarray}
are the traces of tensors $R_{ij}$ and $M_{ij}$.
Furthermore,
${\cal L}_{ij}^\sigma$ 
and ${\cal N}_{ij}^\sigma$ are linear and nonlinear terms, respectively, 
for $\sigma=R$ or $M$, and are given by
\begin{eqnarray}
{\cal L}_{ij}^{\rm R}\!\! &=& \!\!
R_{ik}\overline{U}_{j,k} \! + \! R_{jk}\overline{U}_{i,k}
\!+\!2\Omega_k(\varepsilon_{jkl}R_{il} \! + \! \varepsilon_{ikl}R_{jl}), \\
{\cal L}_{ij}^{\rm M} \!\!\! &=& \!
M_{ik}\overline{U}_{j,k} \! + \! M_{jk}\overline{U}_{i,k}, \\
{\cal N}_{ij}^{\rm R} \!\!\! &=& \!
\!\left(C_1 \! + \! C_4B^2\right)R_{ij}
-C_3 B M_{ij}, \\
{\cal N}_{ij}^{\rm M}\!\!\!\! &=&  \! C_4 B^2 R_{ij}
- (C_3+C_5) B M_{ij},
\end{eqnarray}
where $B=(M/R)^{1/2}$ is the ratio of rms magnetic and velocity fields,
\begin{eqnarray}
{\cal I}_{ij}=C_2(R_{ij} - \onethird R \delta_{ij})
\end{eqnarray}
is an isotropization term, and $C_1 \ldots C_5$ are positive constants 
that are of the order of unity. Advection operators of the form 
$\overline{U}_k \partial_k$ have been neglected. The contribution of 
the advection terms vanish under fully periodic boundary conditions on 
average so they make no contribution. Note also that $\tau$ and $B$ are 
time-dependent, because $R$ and $M$ are time-dependent.

A similar model for the hydrodynamic case has recently been used to to 
model the generation of shear in rotating anisotropic turbulence by the 
$\Lambda$ effect; see K\"apyl\"a \& Brandenburg (2008) for details. An 
important difference is that in their model the flow was driven by an 
external body force, so the equations for the evolution of the Reynolds 
stress have a corresponding forcing term as well. Such a term is here 
absent, because the turbulence is solely the result of shear flow 
instabilities and are modelled by the equations of O03 without external 
forcing. As explained by O03, the model also predicts turbulence for 
$q<0$, where simulations have not shown self-excited turbulence. The 
reason for this is that O03 do not specifically model the MRI dynamics.

In the present study the large scale velocity is given by the shear flow 
$\overline{\bm U} = \bm{U}_0 = -q\Omega_0 x \hat{\bm{e}}_y$. We abbreviate 
the parameter combination $C_i/L$ of O03 by $c_i$. O03 gave an analytic 
solution for the hydrodynamic case. Here we give a partial solution for 
the components of the two stresses if $R$ and $M$ are known. In the steady 
state, the equations yield for the Reynolds stresses
\begin{equation}
R_{xx}=R A \left( \onethird c_2 + Q \right), \label{eq:Rxx}
\end{equation}
\begin{equation}
R_{yy}=R A \left( c_1 + \onethird c_2 + a B^2 c_5- Q\right),
\end{equation}
\begin{equation}
R_{zz}=\onethird R A \, c_2,
\end{equation}
\begin{equation}
R_{xy}=\frac{R^{1/2}}{2q\Omega} \left[c_1 R - (B c_3-c_4) M \right].
\end{equation}
where we have introduced the abbreviations
\begin{equation}
A = \left(c_1 + c_2 + a B^2 c_5\right)^{-1},\quad
a = \frac{c_4}{c_3 + c_5},
\end{equation}
and
\begin{equation}
Q = \frac{2}{q} \left[ c_1 - B^2 \left(B c_3 - c_4 \right)\right].
\end{equation}
The components of the Maxwell stress tensor can be expresses in terms
of the corresponding components of the Reynolds stress tensor as
\begin{equation}
M_{xx}=a B R_{xx}\; ,
\end{equation}
\begin{equation}
M_{yy}=a B \left(R_{yy}-R\right) + M,
\end{equation}
\begin{equation}
M_{zz}=a B R_{zz}\;,
\end{equation}
\begin{equation}
M_{xy}=\frac{R^{1/2}}{2q \Omega} \left[ c_4 - (c_3+c_5)B \right] M. 
\label{eq:Mxy}
\end{equation}
The remaining mixed components involving $z$ vanish, i.e.\
$R_{xz}=R_{yz} = M_{xz}= M_{yz} = 0$.
When $M=0$, there is an explicit expression for $R$, derived by O03.

For the general case with $M\neq0$, we did not find an analytic
expressions for $R$ and $M$ in closed form.
However, using the time-dependent equations it is possible to determine
a linear fit of the form
\begin{equation}
R=R^{(0)}+\sum_{i=1}^5 r_i\left(c_i-c_i^{(0)}\right),
\label{eq:fitri}
\end{equation}
\begin{equation}
M=M^{(0)}+\sum_{i=1}^5 m_i\left(c_i-c_i^{(0)}\right),
\label{eq:fitmi}
\end{equation}
where the $c_{1-5}^{(0)}$ are an approximation to the final fit parameters,
and $R^{(0)}$ and $M^{(0)}$ are the corresponding numerically determined
values of $R$ and $M$. We are thus able to decrease the degree of freedom 
of the fitted system.
We refine the parameters $c_i$ with respect to the initial ``guess'' 
$c_i^{(0)}$ based on the quantity
\begin{equation}
\delta=\left[ (M_{\rm O03}-M_{\rm sim})^2 + (R_{\rm O03}-R_{\rm sim})^2 \right]^{1/2}.
\end{equation}
In the analysis we seek the minimum value of $\delta$.

\section{Simulations}
\label{themod}

For the simulations we adopt a cubic computational domain of size $(2 
\pi)^3$. The gas is isothermal with constant sound speed $c_{\rm s}$; 
vertical gravity and stratification are omitted. The calculations are 
local and therefore made under the shearing box approximation (see below). 
The equations to be solved read
\begin{eqnarray}
\frac{\mathcal{D} \rho}{\mathcal{D} t} &=& - \bm{\nabla} \cdot (\rho \bm{u})\;, \\ \label{cont}
\frac{\mathcal{D} \bm{u}}{\mathcal{D} t}&=&-(\bm{u} \cdot \bm{\nabla})\bm{u} - q \Omega_0 u_x \hat{\bm{e}}_y - 2\,\Omega_0\hat{\bm{e}}_z \times \bm{u} \nonumber \\ && \hspace{0cm}-\frac{1}{\rho}\bm{\nabla} p + \frac{1}{\rho}\bm{J} \times \bm{B} + \frac{1}{\rho} \bm{\nabla} \cdot(2\nu\rho \mbox{\boldmath ${\sf S}$}), \\ \label{velocity}
\frac{\mathcal{D} \bm{A}}{\mathcal{D} t} &=& \bm{u} \times \bm{B} 
    + q \Omega_0 A_y \hat{\bm{e}}_x + \eta \nabla^2 \bm{A}\;, \label{eq:indlocal}
\end{eqnarray}
where $\mathcal{D} / \mathcal{D} t = \partial / \partial t +
U_0 \partial / \partial y$ includes the advection by the shear flow,
$\bm{u}$ is the departure from the mean flow $\bm{U}_0$, $\rho$ is the
density, $\bm{A}$ is the magnetic vector potential, $\bm{B} =
\bm{\nabla} \times {\bm A}$ is the magnetic field, and ${\bm J} =
\bm{\nabla} \times {\bm B}/\mu_0$ is the current density,
$\mu_0$ is the vacuum permeability,
${\sf S}_{ij}=\frac{1}{2}(u_{i,j}+u_{j,i})-\frac{1}{3} \delta_{ij}
\bm{\nabla} \cdot \bm{u}$ is the traceless rate of strain tensor,
$\nu$ is the kinematic viscosity, and $\eta$ is the magnetic diffusivity.

In order to get as close as possible to the ideal limit, we replace the 
diffusion terms by a hyperviscosity scheme, i.e.\ we replace the 
$\nabla^2$-operators by $\nabla^6$, aiming at maximizing the Reynolds 
number in the quiescent regions of the flow while diffusing and damping 
fluctuations near the grid scale. Compared to direct simulations (e.g.\ 
Haugen \& Brandenburg 2006) with uniform viscosities, smaller grid 
resolution can be used to resolve the flow, which is crucial if the plan 
is to, e.g., undertake a parameter study, like in our case.

Periodic boundary conditions are applied in all three directions; in
the radial direction we account for the shear flow $U_0$ by making use
of the shearing box approximation (e.g. Wisdom \& Tremaine 1988):
\EQ
f(\onehalf L_x,y,z)=f(-\onehalf L_x,y+q \Omega_0 L_x t,z),
\EN
where $f$ stands for any of the seven independent variables, $L_x$ stands 
for the radial extent of the computational domain, and $t$ is the time.

The domain is initially threaded by a weak magnetic field,
\EQ
{\bm A} = A_0 \hat{\bm{e}}_y \cos kx \cos ky \cos kz\;,
\EN
where $k=k_1$ has been chosen, and $k_1=2\pi/L_z$ is the smallest
finite vertical wavenumber in the domain of height $L_z$.
Thus, the magnetic field contains periodic $x$- and
$z$-components with amplitude $A_0$.

We choose the values of $k_{\rm A}$, $\Omega_0$ and $A_0$ so that the most 
unstable mode of the MRI, $k_{\rm max}=\Omega /u_{\rm A}$ is well resolved 
by the grid; in practise this means that we always adopt $k/k_1=1$, 
$\Omega_0 = 0.2 c_{\rm s}k_1$ and $A_0 = 0.2 \sqrt{\mu_0\rho_0}\,c_{\rm 
s}k_1^{-1}$, resulting in $k_{\rm max} = O(k_1)$. For the initial setup, 
the other condition for the onset of MRI, namely $\beta \gg 1$, where 
$\beta$ is the ratio of the thermal to magnetic pressure, is also 
satisfied as $\beta$ is minimally 50 at the maximum values of the magnetic 
field. In all our runs the Mach number is well below unity, so 
compressibility effects are negligible.

For all the simulations we use the {\sc 
Pencil-Code}\footnote{\textsf{http://www.nordita.org/software/pencil-code/}}, 
which is a high-order (sixth order in space, third order in time), 
finite-difference code for solving the MHD equations (Brandenburg \& 
Dobler 2002).

The local calculations have been carried out at two different resolutions, 
namely $128$ (in 1D) and $64^3$ (in 3D); the corresponding numerical diffusion 
coefficients are $\nu_{\rm hyper} = \eta_{\rm hyper} = 3.5 \times 10^{-6}$ 
and $2.0\times 10^{-7}$, respectively. The calculations were carried out 
on the IBM eServer Cluster 1600 at Scientific Computing Ltd., Espoo, Finland.

\section{Results}
\label{results}
\subsection{Linear results}

\begin{figure}
\includegraphics[width=\columnwidth]{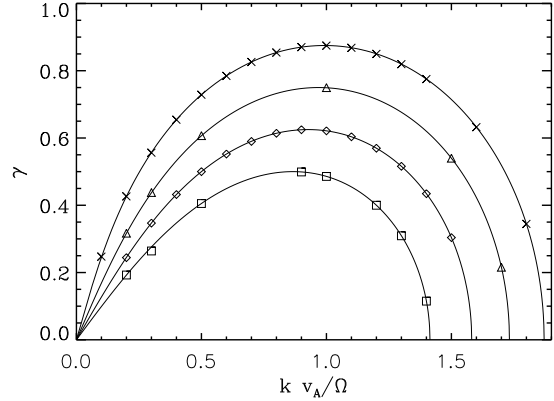}
\caption{Growth rates as a function of wavenumber from the 1D calculations. 
Crosses: $q=1.75$, triangles: $q=1.5$, diamonds: $q=1.25$, squares $q=1.0$. 
The solid lines represent the linear growth rates; see, e.g., PCP06.
}\label{fig:pgrowthrates}
\end{figure}

\begin{figure}
\includegraphics[width=\columnwidth]{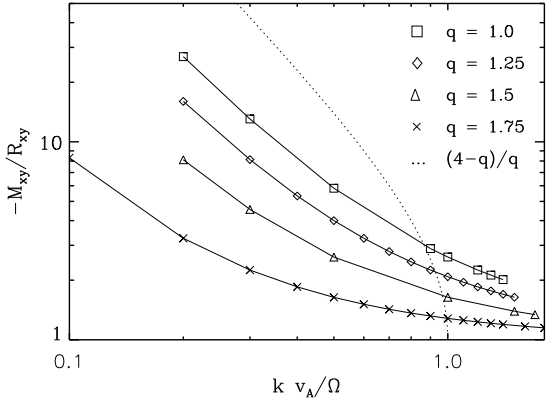}
\caption{Stress ratio $-M_{xy}/R_{xy}$ as function of wavenumber. The solid
lines represent the linear 1D results with varying shear
parameter $q$. The dotted line shows the linear prediction of PCP06
plotted for each $k_{\rm max}(q)$.
}
\label{fig:kstressratios}
\end{figure}

In order to make contact with the model of PCP06, we first aim at 
reproducing their theoretical linear results (see also Balbus \& Hawley 
1991) using one-dimensional calculations with an imposed magnetic field. 
We calculate several sets by fixing the shear rate 
($q=1.00,1.25,1.50,1.75$) and angular velocity $\Omega_0$ and varying the 
initial magnetic field strength $B_0$. The growth rate and the stress 
ratio are monitored during the exponential growth of the instability. The 
results are displayed in Figs.~\ref{fig:pgrowthrates} and 
\ref{fig:kstressratios}.

As can be seen from \Fig{fig:pgrowthrates}, the linear growth rates can be 
reproduced by the numerical method quite accurately. From 
\Fig{fig:kstressratios} it can be observed that the linear prediction 
(dotted curve) intercepts the numerical results (solid lines) exactly at 
$k_{\max}$ for each $q$-curve. At each $k_{\max}(q)$, therefore, the 
linear prediction and numerical results show perfect agreement; see 
Eq.~(\ref{eq:kstress}). In the 1D calculations, however, the system has 
not much freedom to create any other MRI mode than the one that is 
imprinted by the initial magnetic field strength, due to which the stress 
ratio is observed to vary as function of wavenumber, so that monotonically 
decreasing stress ratios are found with increasing wavenumber for all the 
$q$s investigated. This is clearly in disagreement with the PCP06 
assumption, according to which the mode with $k_{\max}$ should always get 
preferentially excited. The wavenumber dependence of the stress ratios, 
however, disappears in 3D: independent of the initial magnetic field 
strength, the mode with $k_{\max}$ is observed to dominate. In that sense 
our 3D results are giving support to the basic assumption of PCP06, 
although, as will be discussed in the remaining part of the paper, the 
magnitudes and $q$-dependence of the stresses are otherwise different from 
the linear analysis.

\subsection{Nonlinear results}

We have performed a set of local calculations, in which we have fixed the 
angular velocity $\Omega_0$ and the strength of the initial magnetic 
field, and then varied the shear parameter $q$. For the investigated range 
of shear parameters $0.4\leq q\leq1.9$, the maximally growing wavenumber, 
therefore, is varying according to Eq.~(\ref{eq:kmax}), but is resolved by 
the numerical grid in all the calculations. We have checked that the 
resulting stresses are independent of the initial magnetic field strength 
(computations with initial plasma $\beta$ of 20--800 were performed). This 
is particularly important in the net flux case (Blackman, Penna \& Varniere 2008).

For smaller shear the stabilizing effect of vorticity is strong and the 
fluctuations grow larger than the stresses themselves; these calculations, 
therefore, are not included in the results. The data averaged over the 
nonlinear saturated stage (typically from a hundred to a few hundreds of 
rotations) for the range $0.4\leq q\leq1.9$ is presented in 
Table~\ref{tab:slabdata}.

\begin{figure*}
\includegraphics[width=0.85\textwidth]{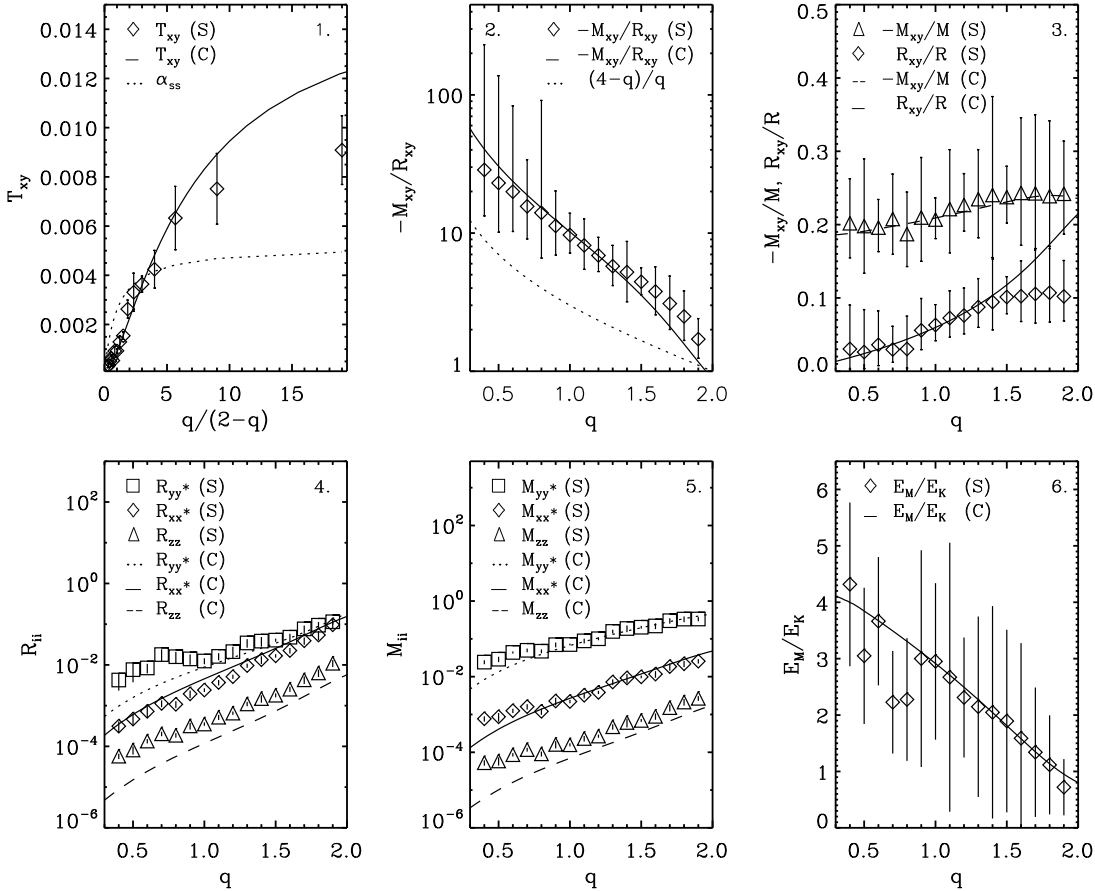}
\caption{Panel 1: total stress as a function of the 
shear-to-vorticity ratio, $q/(2-q)$, from the simulations (diamonds) and 
prediction from the O03 closure model (line) using $c_1 = 0.63$, $c_2 = 
0.73$, $c_3 = 0.33$, $c_4 = 0.58$ and $c_5 = 1.35$. The error bars show 
the standard deviation of the quantity in question. Behaviour 
predicted by the Shakura-Sunyaev $\alpha$ viscosity model has been 
presented for comparison. Panel 2: stress ratio 
as a function of $q$ from the simulations (diamonds), overplotted with the 
O03 closure result (solid line) and the PCP06 linear prediction (dotted 
line). Panels 3-5 show several stress component ratios from the 
simulations overplotted with the O03 closure results. To help 
visualisation, stress components $R_{xx}$, $M_{xx}$ have been scaled up 
with a factor of 6. and $R_{yy}$, $M_{yy}$ with a factor of 20. In panel 
6, the ratio of magnetic to kinetic energy is presented. Throughout the 
figure, (s) is used to denote a simulation result and (c) a result given 
by the closure model.}
\label{fig:cloratio}
\end{figure*}

As is evident from Table~\ref{tab:slabdata}, both Reynolds and Maxwell 
stresses grow with the shear parameter $q$. The growth of the Reynolds 
stress is much stronger than the growth of the Maxwell stress, due to 
which the stress ratio (plotted in the second panel of 
Fig.~\ref{fig:cloratio}) decreases as function of $q$. The simulated 
stress ratio significantly differs from the PCP06 linear prediction given 
by Eq.~(\ref{eq:linstress}) and plotted in the figure with dotted lines. 
All the data points including the error bars are consistently larger than 
the prediction by a factor of 2--3.

Panel 1 of Fig.~\ref{fig:cloratio} shows the total stress, defined in 
Eq.~(\ref{Txy}), as a function of the shear-to-vorticity ratio, $q/(2-q)$. 
The shear-to-vorticity ratio is a quantity that is independent of the 
coordinate system, which was the main reason why Abramowicz et al. (1996) 
presented the stress as a function of this ratio. Interestingly, they 
found that the stress is a nearly linear function of the 
shear-to-vorticity ratio. This is confirmed by the new data. If one were 
to plot the stress as a function of $q$ directly, the relation would 
become strongly nonlinear.

Also the magnetic to kinetic energy ratio (panel 6 in 
Fig.~\ref{fig:cloratio}) exhibits $q$ dependency, but it is less strong 
than that of the magnetic to kinetic stress ratio (panel 2 in 
Fig.~\ref{fig:cloratio}). The energies are defined as $E_{\bm K} = 
\frac{1}{2}\langle \rho \mathbf{u}^2\rangle$ and $E_{\bm M} = 
\frac{1}{2}\langle \mathbf{B}^2\rangle$. Thus $E_{\bm K} = \frac{1}{2}R$ 
and $E_{\bm M} = \frac{1}{2}M$. Our results indicate that for flat 
(galactic) rotation curves with $q=1$ the energy ratio should be a factor of
two higher than for the case of Keplerian accretion disks ($q=1.5$).

\begin{table*}
\caption{Stress components averaged over the saturated regime of the local calculations.
For the O03 closure model, $R_{xz} = R_{yz} = M_{xz} = M_{yz} = 0$.
}
\label{tab:slabdata}
\begin{tabular}{@{}c@{\quad}c@{\quad}c@{\quad}c@{\quad}c@{\quad}c@{\quad}c@{\quad}c@{\quad}c@{\quad}c@{\quad}}
\hline
$q$&$R_{xx}$&$R_{xy}$&$R_{yy}$&$R_{zz}$& $M_{xx}$&$-M_{xy}$&$M_{yy}$&$M_{zz}$&$M/R$\\
\hline
0.4  &$5.3\;10^{-5}$ &$9.7\;10^{-6}$ &$2.1\;10^{-4}$ &$5.7\;10^{-5}$ &$1.3\;10^{-4}$ &$2.8\;10^{-4}$&$1.2\;10^{-3}$ &$5.1\;10^{-5}$ &4.4 \\
0.5  &$7.8\;10^{-5}$ &$1.4\;10^{-5}$ &$3.8\;10^{-4}$ &$8.2\;10^{-5}$ &$1.5\;10^{-4}$ &$3.3\;10^{-4}$&$1.5\;10^{-3}$ &$5.9\;10^{-5}$ &3.1  \\
0.6  &$1.2\;10^{-4}$ &$2.4\;10^{-5}$ &$4.1\;10^{-4}$ &$1.3\;10^{-4}$ &$2.1\;10^{-4}$ &$4.8\;10^{-4}$&$2.1\;10^{-3}$ &$8.4\;10^{-5}$ &3.6  \\
0.7  &$1.9\;10^{-4}$ &$3.8\;10^{-5}$ &$9.0\;10^{-4}$ &$2.0\;10^{-4}$ &$2.7\;10^{-4}$ &$6.0\;10^{-4}$&$2.5\;10^{-3}$ &$1.2\;10^{-4}$ &2.2  \\
0.8  &$2.0\;10^{-4}$ &$3.5\;10^{-5}$ &$8.9\;10^{-4}$ &$1.9\;10^{-4}$ &$2.1\;10^{-4}$ &$5.0\;10^{-4}$&$2.3\;10^{-3}$ &$9.4\;10^{-5}$ &2.0  \\
0.9  &$3.2\;10^{-4}$ &$7.3\;10^{-5}$ &$7.0\;10^{-4}$ &$3.1\;10^{-4}$ &$3.7\;10^{-4}$ &$8.2\;10^{-4}$&$3.4\;10^{-3}$ &$1.6\;10^{-4}$ &2.0  \\
1.0  &$4.2\;10^{-4}$ &$9.1\;10^{-5}$ &$6.3\;10^{-4}$ &$3.7\;10^{-4}$ &$3.8\;10^{-4}$ &$8.7\;10^{-4}$&$3.6\;10^{-3}$ &$1.7\;10^{-4}$ &2.9  \\
1.1  &$6.6\;10^{-4}$ &$1.6\;10^{-4}$ &$8.5\;10^{-4}$ &$5.6\;10^{-4}$ &$6.0\;10^{-4}$ &$1.3\;10^{-3}$&$4.8\;10^{-3}$ &$2.6\;10^{-4}$ &2.7 \\
1.2  &$8.6\;10^{-4}$ &$2.0\;10^{-4}$ &$1.1\;10^{-3}$ &$6.6\;10^{-4}$ &$6.4\;10^{-4}$ &$1.3\;10^{-3}$&$5.0\;10^{-3}$ &$2.7\;10^{-4}$ &2.3 \\
1.3  &$1.7\;10^{-3}$ &$4.1\;10^{-4}$ &$1.8\;10^{-3}$ &$1.2\;10^{-3}$ &$1.3\;10^{-3}$ &$2.4\;10^{-3}$&$8.4\;10^{-3}$ &$5.2\;10^{-4}$ &2.2   \\
1.4  &$2.5\;10^{-3}$ &$6.1\;10^{-4}$ &$2.2\;10^{-3}$ &$1.7\;10^{-3}$ &$1.8\;10^{-3}$ &$3.1\;10^{-3}$&$1.1\;10^{-2}$ &$7.2\;10^{-4}$ &2.1   \\
1.5  &$2.7\;10^{-3}$ &$6.4\;10^{-4}$ &$2.0\;10^{-3}$ &$1.8\;10^{-3}$ &$1.6\;10^{-3}$ &$2.9\;10^{-3}$&$9.8\;10^{-3}$ &$6.8\;10^{-4}$ &1.9  \\
1.6  &$3.7\;10^{-3}$ &$8.5\;10^{-4}$ &$2.3\;10^{-3}$ &$2.5\;10^{-3}$ &$1.8\;10^{-3}$ &$3.2\;10^{-3}$&$1.0\;10^{-2}$ &$8.4\;10^{-4}$ &1.6   \\
1.7  &$6.3\;10^{-3}$ &$1.5\;10^{-3}$ &$3.7\;10^{-3}$ &$4.1\;10^{-3}$ &$2.9\;10^{-3}$ &$4.6\;10^{-3}$&$1.5\;10^{-2}$ &$1.4\;10^{-3}$ &1.3   \\
1.8  &$8.9\;10^{-3}$ &$2.0\;10^{-3}$ &$4.4\;10^{-3}$ &$6.1\;10^{-3}$ &$3.4\;10^{-3}$ &$5.1\;10^{-3}$&$1.6\;10^{-2}$ &$2.0\;10^{-3}$ &1.1   \\
1.9  &$1.8\;10^{-2}$ &$3.7\;10^{-3}$ &$6.2\;10^{-3}$ &$1.1\;10^{-2}$ &$4.8\;10^{-3}$ &$6.2\;10^{-3}$&$1.8\;10^{-2}$ &$2.9\;10^{-3}$ &0.7   \\
\hline
\end{tabular}
\end{table*}

\subsection{Predicting stresses with the O03 model}

In their treatment, O03 use fiducial closure parameters $C_{1-5}=1$ in 
order to demonstrate the overall behavior of the model. He recommends the 
parameters to be calibrated by comparison with numerical simulations in 
order to obtain more accurate predictions. We have made an attempt to 
determine the dimensionless closure parameters that work for each value of 
$q$ in the range $0.4 \leq q \leq 1.9$.

There are essentially two ways to approach the problem. The first is 
to take the time independent equations (\ref{eq:Rxx})--(\ref{eq:Mxy}) and 
solve for $c_{1-5}$ using values of the traces $R$ and $M$ from the 
simulations. We call this method ``backward modelling''. The problem with 
this method is that the Eqs.~(\ref{eq:Rxx})--(\ref{eq:Mxy}) are incomplete 
and do not set any constraints to the ratio $M/R$. Consequently, it does 
not produce the correct solution to the time dependent equations 
(\ref{eq:timdepeq}) and (\ref{eq:timdepeq2}). Therefore we also used the 
linear approximation described by Eqs.~(\ref{eq:fitri}) and 
(\ref{eq:fitmi}) to improve the fit.

\begin{figure}
\includegraphics[width=\columnwidth]{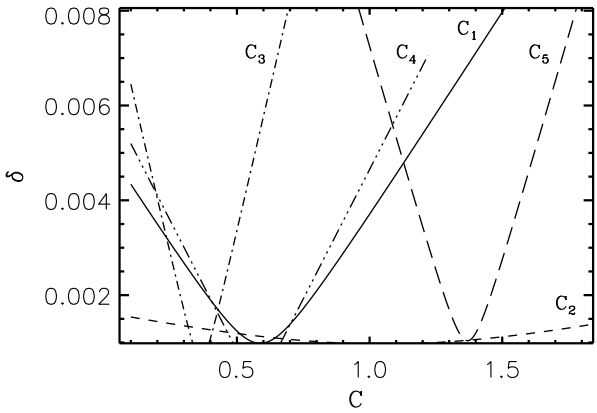}
\caption{
The error estimate $\delta$ (see Eq.~(28)) as a function of $C$.
}
\label{fig:Cdiff}
\end{figure}

The second way of determining $c_{1-5}$ is what we call ``forward 
modelling''. The idea here is to seek such $c_{1-5}$ that the 
results of the time-dependent equations (\ref{eq:timdepeq}) and 
(\ref{eq:timdepeq2}) yield the same individual stresses and the traces $R$ 
and $M$ as the numerical simulations. This time a universal $c_{1-5}$ is 
determined so that it predicts the stresses for all values $0.4 \leq q 
\leq 1.9$. Once a reasonably good set of $c_{1-5}$ was found, the result 
was fine tuned further using the linear approximation given by 
Eqs.~(\ref{eq:fitri}) and (\ref{eq:fitmi}); see 
\Figs{fig:Cdiff}{fig:closure}. From the \Fig{fig:Cdiff} it can be seen 
that the best fit is obtained by fixing $c_4$ while keeping the other 
$c_i$ unchanged.

The final closure parameters are thus $c_1 = 0.63$, $c_2 = 0.73$, $c_3 = 
0.33$, $c_4 = 0.58$ and $c_5 = 1.35$, corresponding to $C_i = c_i L$ with 
$C_1 = 4.0$, $C_2 = 4.6$, $C_3 = 2.1$, $C_4 = 3.6$ and $C_5 = 8.5$.
For these $c_{1-5}$ the relevant fit parameters in 
Eqs.~(\ref{eq:fitri}) and (\ref{eq:fitmi}) are $R^{(0)}=0.0057$ and $M^{(0)}=0.013$ 
together with $r_1=-0.0065$, $r_2=-0.00012$, $r_3=0.024$, $r_4=-0.010$, 
$r_5=-0.0061$, and $m_1=-0.0057$, $m_2=-0.0012$, $m_3=0.0042$, 
$m_4=0.0038$, $m_5=-0.018$. 

\begin{figure}
\includegraphics[width=\columnwidth]{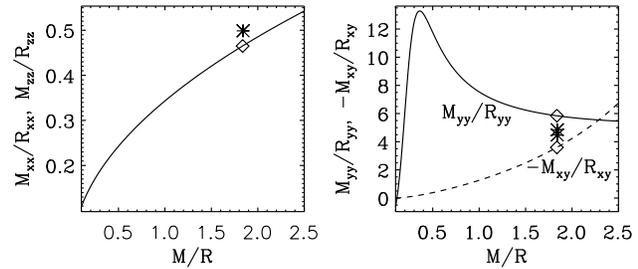}
\caption{
Comparison of the simulation data (diamonds) to the chosen fit to
Ogilvie model with $c_1=0.63$, $c_2=0.73$, $c_3=0.33$, $c_4=0.58$,
$c_5=1.36$ (diamonds) for one particular run with $q$=1.5. The solid
line shows the $M/R$-dependent result of the stationary equations with
the chosen set of $c$-parameters.  The additional constraint used in
the fitting procedure required that the difference in the simulated
and model $M/R$ is minimal, due to which in this solution the values
of $M/R$ match while the individual stress ratios somewhat differ. }
\label{fig:closure}
\end{figure}

\section{Discussion}
\label{discuss}
In this study we set out to investigate the shear rate dependency of 
MRI-generated turbulent stresses. We have performed a series of local 
shearing box simulations with varying $q$ and measured the resulting 
turbulent stresses. We find that the turbulent stress ratio 
$-M_{xy}/R_{xy}$,  and the total stress $T_{xy}$ exhibit strong 
$q$-dependency. 
The relation for the stress ratio by PCP06; see
Eq.~(\ref{eq:linstress}) predicts similar behaviour, but the ratio 
computed from the simulations is consistently 2-3 times larger than what 
their result indicates.

In order to further study the evolution of the MRI and the stresses we 
have attempted to reproduce them using the local closure model by O03. We 
first find a set $c_{1-5}$ such that the time-dependent equations give the 
same individual stresses and traces $R$ and $M$. The linear approximation 
of Eqs.~(\ref{eq:fitri})--(\ref{eq:fitmi}) is then used together with the 
time-independent equations to improve the fit. The O03 closure parameters 
that describe our numerical simulation results are found to be $C_1 = 
4.0$, $C_2 = 4.6$, $C_3 = 2.1$, $C_4 = 3.6$ and $C_5 = 8.5$.

The closure model by O03 was thus found to predict our simulation 
generated turbulent stresses quite well. The model certainly 
offers a much-needed method for studying the evolution of turbulent 
stresses in the shearing sheet limit. However, additional constraints for 
$R$ and $M$ are needed in order to determine the closure parameters $C_i$. 
The linear fitting approximation described in Eqs.~(\ref{eq:fitri}) and 
(\ref{eq:fitmi}) was devised to overcome this shortcoming.

Our results may also have some relevance in the galactic context,
in which the MRI has been proposed to be responsible of the anomalous
turbulent velocity dispersions found in the outer regions of some galaxies
(e.g.\ Sellwood \& Balbus 1999). On the other hand, the energy balance 
estimates from observations of NGC6949 (Beck 2004) indicate that magnetic 
energy could become dominant over the kinetic energy in the outer regions 
of this galaxy, so that the energy ratio $E_{\rm M}/E_{\rm K} \approx 3$--$4$.
According to our present results, in MRI-driven systems the magnetic 
energy clearly dominates over the kinetic energy for all $q<1.6$, at the
galactic value of $q=1$ for flat rotation curves we find the value 
$\approx 3$, a number agreeing rather well with the observational 
estimates.

\acknowledgements{It is a pleasure to thank an anonymous referee for a 
comprehensive report that lead to significant improvements in the 
manuscript. The authors also wish to express their gratitude to 
Nordita for their hospitality during the program `Turbulence and Dynamos.' 
We acknowledge the Scientific Computing Ltd.,
Espoo, Finland, for granting CPU time in their supercomputers.
This work was supported by the Academy of Finland through grants
No.\ 112020 (AJL) and No.\ 121431 (PJK) and in part by the
National Science Foundation under grant PHY05-51164 (AB).}



\end{document}